*The use of diffractive screens for electronic imaging*

Prof. José J. Lunazzi
UNICAMP
Instituto de Física
<lunazzi@digicron.com>
<http://www.digicron.com/lunazzi>

Abstract:

Diffractive screens are high-resolution elements with capability for generating holographic-like images from a sequence of planes where TV frames are seen oblique to it. If we project a sequence of contour lines of an object it could be seen in continuous horizontal parallax to a size of $1m^3$.

___________________________________________________

# THE USE OF DIFFRACTIVE SCREENS FOR ELECTRONIC IMAGING

José J. Lunazzi
Campinas State University
Physics Institute - C.P.6165
13083-100 Campinas-SP-Brazil
lunazzi@ifi.unicamp.br

## 1. Introduction

Diffractive screens are known since long time ago, and its connection to holographic images, although not direct, comes from the inventor of holography, Dennis Gabor. He invented the projection of a stereoscopic pair onto angularly selective diffractive screens made by two-beam interference restricted to monochromatic light.  His invention was followed by some similar work[1] and extensions to white light[2,3].

To our knowledge, the only known demonstrations of the use of diffractive screens for electronic imaging  under  white light should be attributed to Lunazzi[4] who developed a gogglesless stereoscopic TV system in 1988 .  He made public demonstrations of it beginig 1989, including registering and simultaneous transmission[5,6,7].

The basic scheme for this system is shown in Fig. 1.

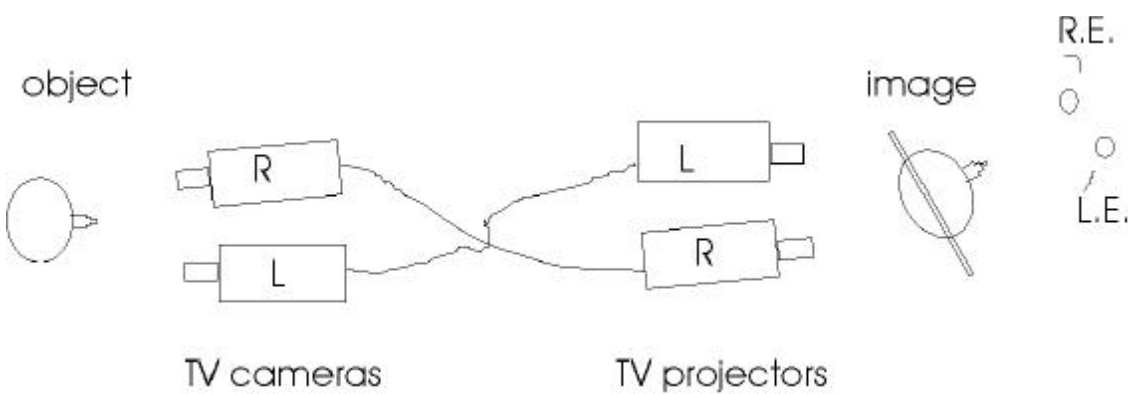

**Fig.1**

The pair of cameras and projectors can be replaced by single elements when mirrors acting as beam splitters divide the field for the image.  The images can be made and projected under white light to a size of up to 15cm x 15cm.  The angle of diffraction was $30^0$ creating little distortion on the lateral size.  Since most wavelengths are concentrated on the region where the right eye R.E. and the left eye L.E. of the observer are located color reproduction is partially achieved.

The way in which each projected image is viewed from an specific viewpoint can be understood from Fig. 2, where dashed lines with longer segments represents a longer wavelength than for the case of lines with smaller segments.

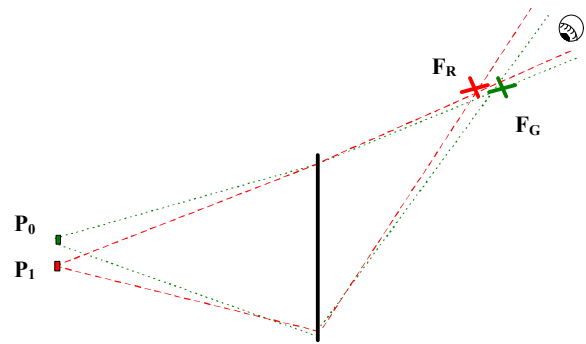

**Fig. 2**

The region allowed for the observer is rather unique, and because is a stereo system, stereo clues can not be solved. But it is an interesting system that indicated the way to a more complete one satisfying this requirements.

Very few possibilities are referred to give a solution to the problem of holographic television. We mean here holographic being any system capable of bringing images in continuous horizontal parallax. Occlusion is a necessary property for representing objects with realm, as opaque elements instead of transparent ones.

The only system we know that satisfies this requirements is being developed by Benton[8] since 1989. It employs very powerful parallel computing processing, crystal electrooptical modulators, acoustic waves and lasers. Color can be obtained by making a three-chromatic system. The computer simulated images are, to our knowledge, well defined and in continuous horizontal parallax . It was recently reported[9] to have reached the size 8.5cm (V) x 13cm (H) x 20 cm$^3$. Good microcomputers can be used for exhibiting line rigid objects ("wire frames") in rotation[10].

We proposed a white light system capable of generating a vectorially addressed holo-like image [11,12,13]. This system was further developed recently to generate a sequence of TV planes where reach TV frame is seen oblique to a holographic screen, traversing it from its front to his back. A controlled mirror makes this plane to fill a volume by scanning along the screen, so that if we project a sequence of contour lines of an object it can be seen in continuous horizontal parallax, to a size of 1m$^3$, up to now. First results using computer generated models will be described. The system has the possibility of occlusion by control of the spectral distribution when encoding each point, a development to reach in future work.

## 2. Description

A diffractive directional screen was described by us[14] as a diffractive lens where images are projected. We see in Fig. 3 how it can distribute different information according to wavelength

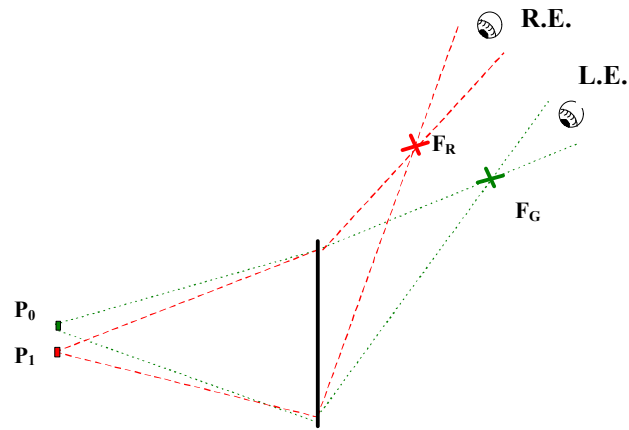

**Fig. 3**

An image in pure red, for example, can be seen by the right eye **R.E.** while an entirely different image in pure green can be seen by the left eye **L.E.**. **P₀** and **P₁** are points on a projecting lens, and may coincide. Following this example we can see in Fig. 4 how the presence of a point in three dimensional space can be established by having one representative element for each wavelength. One green and one red point can be imaged at a certain horizontal distance on the screen giving a virtual image behind the screen, or can coincide at the same position for giving the image on the screen, or in reversed sequence at a certain distance to converge light in front of the screen.

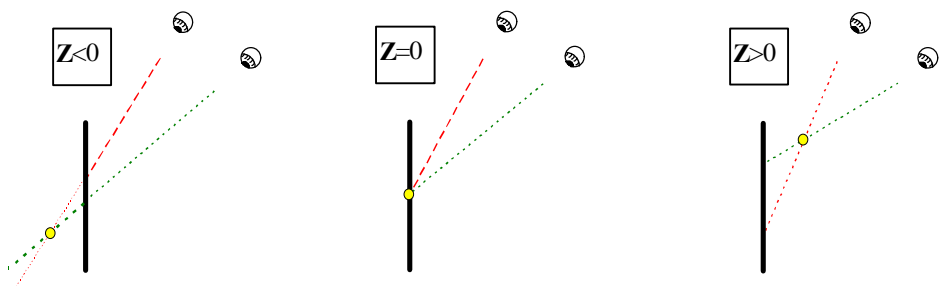

**Fig. 4**

The principle that made evident this possibility to us was the chromatic encoding of views discovered as a natural diffraction phenomena[15] and its reciprocal decoding obtained by a grating[16] or a holographic screen[15,17].

Many experiences gave images in unlimited depth whose appearence was a perfect matching to the properties of holographic images being reproduced under white light[18,19]. The most impressive result was the possibility of enlarging holograms under white light[20,21] from a plastic film on the 35mm format. We show it in Figure 5 for suggesting the possibility of also achieving electronic holo-imaging to this large size (0.75m H x 1.14m V).

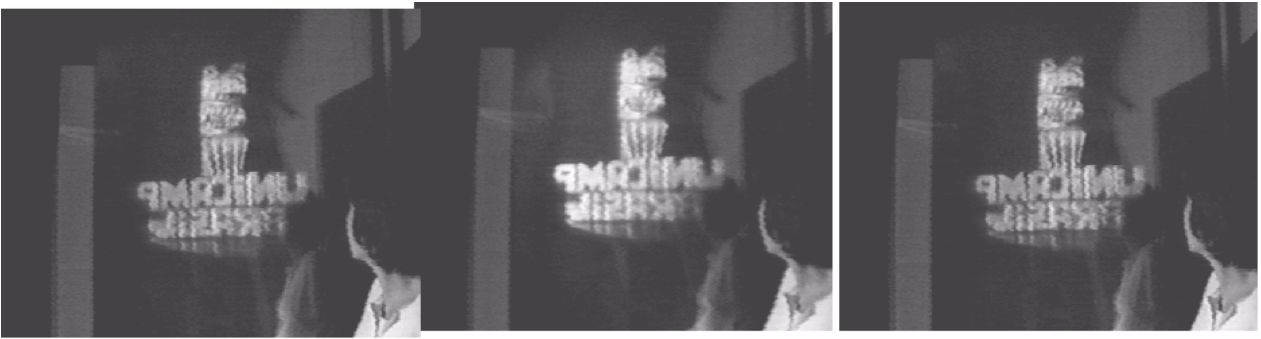

**Fig. 5**

## 3. Experimental Details

We made a screen 0.75m (H) x 1.14m (V) on holographic film where images can be represented without any limit in depth, but usually are 1m deep. Our previous system was vectorially addressed by focusing the white light image of a small lamp onto a grating **G** (see Fig. 6) rendering diffracted images on a diffraction grating, to be focused on the screen at an incident angle of $45^0$.

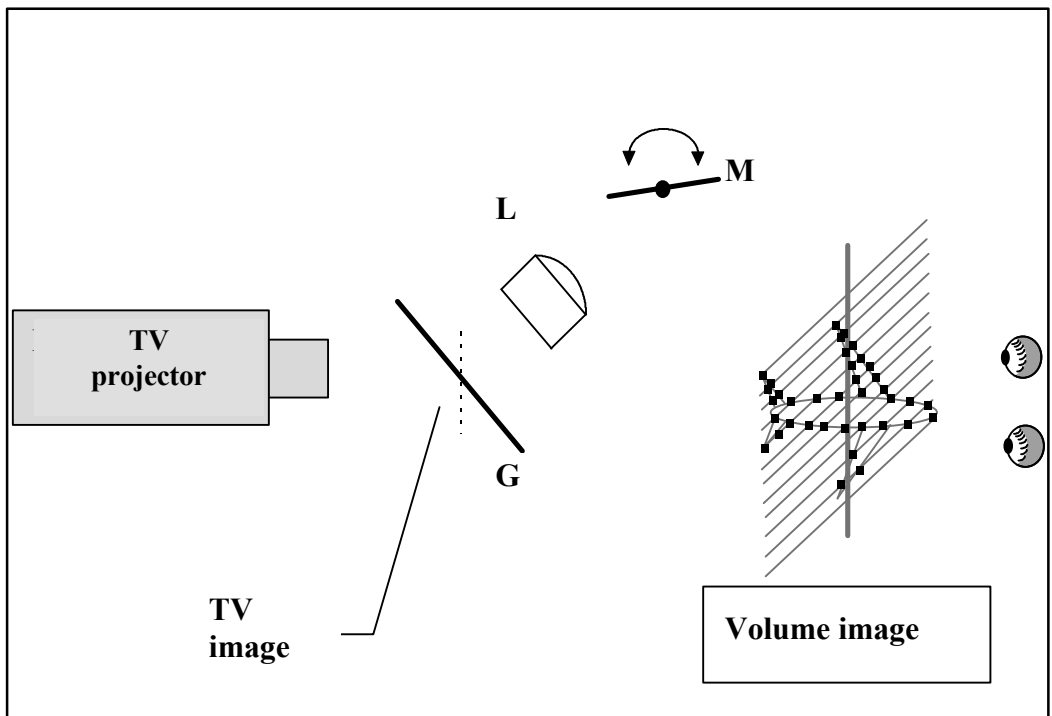

**Fig. 6**

We employed[22] the same setup than for a vector image[13], replacing the point source by a plane TV frame coming from a LCD TV projector. It appears as a dotted line oblique to the grating and each one of the set of lines oblique to the screen can contain one TV frame. By drawing several contour lines the object can be represented. We show the result obtained with computer generated rings as contour lines in Fig.7. It is a stereo pair in the form of left-right-left views to allow for direct observation so that one can choose the first and second image for parallel stereoscopic viewing, or second and third view for crossed viewing . A single ring was drawed, and part of an ellipse. They appear three times due to persistence. The elliptic element does not appears oblique to the screen but parallel in front and wider becasue entered into a region where the system strongly aberrrates.

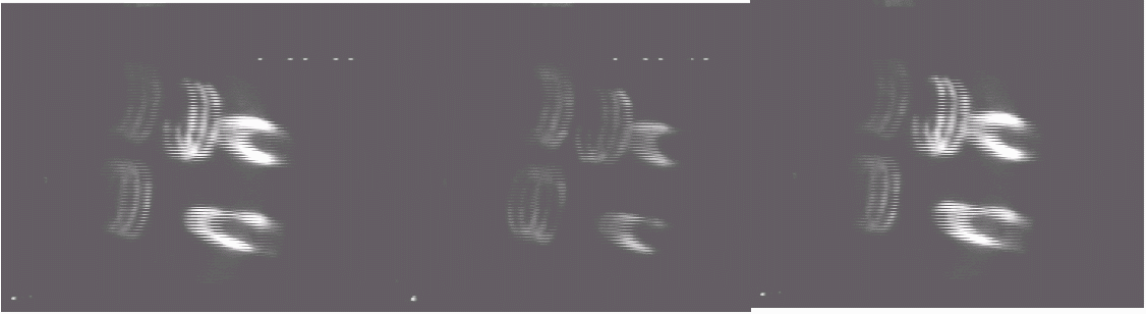

**Fig. 7**

Screen size was 66cm(H) x 38cm(V) and the projector was SHARP XU-400G activated in RGB mode by means of a TARGA +64 card. The three-dimensional appearance of a cylinder becomes evident in this result composed of only three planes. The three planes we show in the photographs is precisely the same amount that an observer sees due to persistence of the image. The field of view at a distance of 70cm from the screen was 25cm. We can now easily extend this experience to the screen size of 0.75m (H) x 1.14m (V) already employed for another related experiences [7].

## 4. Conclusions

We demonstrated the possibility of a new system for TV images having continuous horizontal parallax, same as holographic images. Large size, simple lighting and simple computer requirements are their qualities. It is presented in its initial form and much work is still to be done to reach the capability of simultaneous registering, transmission and reception. It may constitute a way to overcome the main limitations of present goggless stereoscopic systems which are restricted to limited depth and, when allows for freedom of movement of the observer, only one observer is allowed.

## 5. Acknowledgements

The Foundation for Assistance to Research of the Sao Paulo State-FAPESP and "Financiadora de Estudos e Projetos"-FINEP are gratefully acknowledged.